\documentclass[12pt]{iopart}

\usepackage{graphicx,amssymb,bbm}

\def\tr{\,{\rm tr}\,}
\def\ket#1{|#1\rangle}
\def\bra#1{\langle#1|}
\def\braket#1#2{\langle #1 | #2 \rangle}

\def\ii{{\rm i}}
\def\vec#1{\underline{#1}}
\newcommand{\mm}[1]{{\mathbf{#1}}}

\begin{document}

\title[Simulations of non-equilibrium steady states]{Matrix product simulations of non-equilibrium steady states of 
quantum spin chains}
\author{Toma\v z Prosen and Marko \v Znidari\v c}
\address{Physics Department, Faculty of Mathematics and Physics,
University of Ljubljana, Jadranska 19, SI-1000 Ljubljana, Slovenia}

\begin{abstract}
Time-dependent density matrix renormalization group method
with a matrix product ansatz is employed for explicit computation of 
non-equilibrium steady state density operators of several integrable and
non-integrable quantum spin chains, which are driven far from equilibrium
by means of Markovian couplings to external baths at the two ends.
It is argued that even though the time-evolution can not be simulated efficiently
due to fast entanglement growth, the steady states in and out of equilibrium
can be typically accurately approximated, so that chains of length
of the order $n\approx 100$ are accessible. Our results are demonstrated
by performing explicit simulations of steady states and calculations of 
energy/spin densities/currents in several problems of heat and spin transport in quantum spin chains.
Previously conjectured relation between quantum chaos and normal transport is 
re-confirmed with high acurracy on much larger systems.
\end{abstract}

\submitto{{\it J.~Stat.~Mech.}}
 
\section{Introduction}

A detailed understanding of the properties of strongly interacting many-particle quantum systems represents one of the major challenges of
theoretical physics. Even for one spatial dimension (1D) and local interactions very little is known about physics at high temperature, out of equilibrium, or for long real time evolution.

Quite recently, several original ideas emerging from the quantum information theory on the issue of entanglement in many-particle systems \cite{fazioreview} resulted in very useful efficient techniques for the ``classical" simulation of quantum many particle systems \cite{vidal,ciracreview}.  Even though most versions of these methods can be re-interpreted in terms of the {\em density matrix renormalization group} \cite{white} (for an excellent review see \cite{ulireview}) and its time-dependent generalizations \cite{uli,whitetime} they bring up a conceptual simplicity: namely, they represent the quantum many body states in terms of the so-called finitely correlated states \cite{werner} which can be written in terms of matrix products, the so-called {\em matrix product states} (MPS). Similarly, {\em matrix product operator} (MPO) ansatz can be used to describe density operators of mixed states \cite{zwolak,verstraete}.

Yet, most of these ingenious numerical methods turned out to be effective {\em only} for calculation of ground states, low temperature equilibrium states, or time evolved states for short time or starting from particular (e.g. few ``quasi-particle") initial conditions. The calculation of asymptotic time evolution in the thermodynamic limit, and the properties of {\em non-equilibrium steady states} (NESS), remains notoriously difficult due to typical growth of bi-partite entanglement in generic non-integrable systems and consequent exponential growth of the dimension of the relevant many-particle Hilbert space \cite{PRE:07}. 
Therefore, the regularity to quantum chaos transition is manifested in the difficulty of 
classical simulation of quantum many-body dynamics \cite{PRE:07}, see also \cite{andreas}.
Even though promising analytical techniques based on Fock spaces of operators have been developed recently in terms of which one can access NESS and relaxation properties of certain open integrable quantum many-body systems far from equilibrium \cite{NJP:08,arxiv:08}, for generic non-integrable systems one still needs to rely on more or less sophisticated ``brute force" methods limited at present to 20-30 spins 1/2 (or qubits, fermions) in 1D. For such small systems it is often difficult or impossible to conclude on asymptotic thermodynamic properties due to considerable finite size effects. 

In this paper we shall put forward a hypothesis that NESS of certain {\em open} 1D systems can be efficiently simulated in terms of MPO ansatz.

Our idea is based on the following arguments.
It has been shown rigorously that ground states of several interacting many-particles systems in 1D can be written in terms of MPS (for example, the so-called {\em valence bond states} \cite{afleck}). Later, it has been shown \cite{calabrese,kitaev}, using the arguments of conformal field theory, that ground state entanglement entropy of a finite block of size $n$ in any non-critical (gaped) system in 1D 
saturates with $n$, whereas for critical (non-gaped) systems the growth of entanglement entropy 
is at most logarithmic in $n$.
This result immediately implies that an efficient MPS representation of ground states in 1D should exist with the dimension of the 
parameterizing matrices which saturates or grows not faster than polynomially with the system 
size $n$. Similar ``strong efficiency" result can be claimed for thermal states \cite{arxiv:08b}. The defining equation for NESS of an open-quantum system can be written as ``right ground state" of a 
certain non-Hermitian quantum Liouville master operator \cite{NJP:08}, hence NESS can be thought of as a kind of non-normal ground state in the Liouville space. This means that the entanglement of NESS, treated as an element of the Hilbert space of operators, can be relatively weak for a
a wide class of (non-integrable) problems even though the operator space entanglement of generic operators is much stronger (e.g. of the ``excited'' eigenstates of Liouville master operator, i.e. decay modes). 
If this intuition is correct then NESS should be simulable in terms of MPO ansatz for systems considerably longer than $30$ spins (at present).

In the present work we are going to test MPO method on NESS calculation of different 1D systems, integrable and chaotic, as well as normal and ideal conductors. In particular, we shall focus on the problem of diffusive versus ballistic spin/heat transport in spin chains. There have been basically two approaches to quantum transport. The first one is using the linear response formalism, calculating equilibrium time correlation functions. In that approach one studies purely Hamiltonian system, {\em i.e.}, without any external baths, either directly (see e.g.~\cite{Castella:95,Narozhny:98,Zotos:99,Saito:03,Brenig:03,Long:03,Sakai:03,Pereira:06}), or in terms of conformal field theory~\cite{Fujimoto:03}, or quantum Monte Carlo~\cite{Alvarez:02}. For exact calculations spin chains of sizes up to $n=28$ are achievable using a micro-canonical finite temperature Lanczos method~\cite{Long:03}. For a discussion of applicability of Green-Kubo formalism see~\cite{Gemmer:06}. The second approach goes via direct simulation of out-of-equilibrium system coupled to the baths, usually using a master equation~\cite{Saito:02,Michel:03,Mejia:05,Steinigeweg:06,Mejia:07,Wichterich:07,Yan:08,Michel:08} solving it by either numerical integration, diagonalization, or Monte Carlo wave-function approach. Using these methods one can study chains of up to $n=20$ spins~\cite{Mejia:07,Michel:08}. 

Using MPO approach for solving master equation proposed in the present paper we can calculate NESS of $\sim 100$ spins. Even though we are at present unable to give a precise statement about the efficiency of the method for a general NESS (for an exception due to exact solvability see~\cite{arxiv:08}), MPO approach should nevertheless prove very useful in studying out-of-equilibrium quantum many-body phenomena.

In section \ref{sec:master} we outline our numerical algorithm of computation of NESS in terms of MPO ansatz as an asymptotic time-dependent solution of quantum (Lindblad) master equation. We also describe in details efficient models of the baths used in our calculations.
In sections \ref{sec:main1} and \ref{sec:main2} we discuss our numerical results for NESS and their transport properties in 
several integrable and non-integrable models of quantum spin chains, whereas in section \ref{sec:disc} we summarize our findings and conclude.

\section{Master equation}  

\label{sec:master}

The most general completely positive trace preserving flow can be written in a form of the Lindblad equation \cite{lindblad} 
(see \cite{alicki} for a comprehensive review on completely positive flows)
\begin{equation}
\frac{{\rm d}}{{\rm d}t}{\rho}=\ii [ \rho,H ]+ \gamma \sum_k \left( [ L_k \rho,L_k^{\dagger} ]+[ L_k,\rho L_k^{\dagger} ] \right),
\label{eq:Lin}
\end{equation}
(setting $\hbar=1$) where $H$ is the Hamiltonian, generating the unitary part of evolution, $L_k$ are Lindblad operators, and $\gamma$ is some overall bath-coupling strength parameter. Formally, the solution of linear Lindblad equation can be written as $\rho(t)=\exp{( \hat{\cal L}t)}\rho(0)$, where $\hat{\cal L}$ is a Liouvillean super-operator corresponding to the right-hand-side of the Lindblad eq.(\ref{eq:Lin}). 

\subsection{Matrix product operators and simulation of time-evolution}

We are going to simulate the time-evolution under the Lindblad equation by using a matrix product operator (MPO) formulation.
For a chain of $n$ spins 1/2,  the density matrix $\rho$ can always be expanded over all possible products of local Pauli operators which form a basis of $4^n$ dimensional Hilbert space of operators, 
\begin{equation}
\ket{\rho} = \sum_{\vec{s}}  c_{\vec{s}} \ket{\sigma^{\vec{s}}},
\label{eq:superket}
\end{equation}
where we use short notation $\sigma^{\vec{s}}=\sigma_1^{s_1} \cdots \sigma_{n}^{s_{n}}$, $\vec{s}\equiv s_1,\ldots s_n$, and $s_i \in \{0,1,2,3\}$, with $\sigma^0=\mathbbm{1}, \sigma^1=\sigma^{\rm x}, \sigma^2=\sigma^{\rm y}, \sigma^3=\sigma^{\rm z}$. A lower index in Pauli operators, for instance $l$ in $\sigma^s_l$, will always denote a position $l \in \{1,\ldots n\}$ of the spin on which it operates. In the MPO ansatz, the expansion coefficients $c_{\vec{s}}$ are represented in terms of $4n$ $D\times D$ matrices $\mm{A}_i^{s_i}$, $i=1,\ldots,n$, as
\begin{equation}
c_{\vec{s}} = \tr{(\mm{A}_1^{s_1} \cdots \mm{A}_{n}^{s_{n}})}.
\label{eq:MPO}
\end{equation}
The propagator $\exp{(\hat{\cal L}t)}$ is implemented within the MPO formalism in small time steps of length $\tau$ (in our simulations $\tau=0.05$). We decompose Liouvillean as 
$\hat{\cal L} =
\hat{\cal L}_{1} + \hat{\cal L}_{2}$, with the condition that all the terms grouped within each
$\hat{\cal L}_\nu$, $\nu=1,2$, mutually commute. For example, $\hat{\cal L}_{1}$ 
contains terms with interactions between 2nd and 3rd spin, 4th and 5th spin, and so on, 
while $\hat{\cal L}_{2}$ contains interactions between other pairs, whereas the corresponding
one-body terms are distributed evenly between $\hat{\cal L}_{1}$ and $\hat{\cal L}_{2}$. 
For each small time step we use Trotter-Suzuki formula, 
$\exp{(\hat{\cal L}\tau)}= 
\prod_k \exp{(\alpha_k \hat{\cal L}_{1}\tau)}\exp{(\beta_k \hat{\cal L}_{2}\tau)}
+{\cal O}(\tau^p)$, with an appropriate scheme having either $p=3$ or $p=4$, depending on the required accuracy.
MPO representation of a density matrix $\rho$ is exact only if the matrix dimension $D$ of a matrices $\mm{A}_l^s$ at $l$-th site is equal to the number of nonzero Schmidt coefficients $\mu_j$ for a {\em bi-partite} splitting of a super-ket $\ket{\rho}$ to the first $l$ sites, denoted by $A$, and the remaining $n-l$ sites, denoted by $B$,
\begin{equation}
\ket{\rho}=\sum_j \mu_j \ket{\xi^{\rm A}_j} \ket{\xi^{\rm B}_j},
\end{equation}
with orthogonal vectors $\ket{\xi^{\rm A}_j}$ and $\ket{\xi^{\rm B}_j}$. The inner product is defined by $\braket{\alpha}{\beta}=\tr{(\alpha^\dagger \beta)}/2^n$. Since the maximal number of nonzero Schmidt coefficients is usually exponentially large in $n$ one truncates exact representation by keeping only a small fraction of largest $\mu_j$. An error made in such a truncation is given by a sum of truncated $\mu_j^2$ \cite{wolf}. As a rough estimate of a minimal needed matrix dimension $D$ one can use Von Neumann entropy of a super-ket denoted by $S^\sharp$, also called {\em operator space entanglement entropy} (OSEE) (for an earlier definition in a different context see~\cite{operator}). In terms of coefficients $c_{\vec{s}}$ OSEE can be expressed as,
\begin{equation}
S^\sharp=-\tr_A{(\mm{R}\,\log_2{\mm{R}})},\qquad \mm{R}=
\braket{\rho}{\rho}^{-1} \tr_B{ \ket{\rho} \bra{\rho}},
\label{eq:S}
\end{equation}
where, again, a subscript $A$ denotes the first $l$ sites and $B$ its complement. Writing the expansion coefficients $c_{\vec{s}}$ as a $2^l\times 2^{n-l}$ dimensional matrix $\mm{C}_{\vec{s}_A,\vec{s}_B}=c_{\vec{s}_A \vec{s}_B}$, the matrix $\mm{R}$ (\ref{eq:S}) is given by $\mm{R}=\mm{C} \mm{C}^T/\tr{(\mm{C} \mm{C}^T)}$. OSEE is a measure of entanglement of a super-ket $\ket{\rho}$ which, however,
is essentially different quantity as entanglement of a mixed state represented by the density operator $\rho$. Also, OSEE is different than the ordinary von Neumann entropy of $\rho$. 
For instance, OSEE either saturates or grows logarithmically with time in integrable transverse Ising model~\cite{PRA:07} or in Heisenberg XXZ model in a random magnetic field~\cite{PRB:08}, it saturates or grows linearly with $n$ in NESS of XY spin chain~\cite{arxiv:08}, and it saturates or grows logarithmically with the inverse temperature for generic thermal states~\cite{arxiv:08b}. 

\subsection{Physics and implementation of the baths}

Let us say a few words about the details of our implementation of Markovian baths, that is of a non-unitary part $\hat{\cal L}_{\rm B}$ of $\hat{\cal L}$, involving Lindblad operators $L_k$. 
In our case - as our goal is to simulate {\em many-body coherent transport} in one dimension - the baths, i.e., the Lindblad operators $L_k$, will act only at the two {\em ends} of a 1D chain. More precisely, they will act just on the first and the last spin for a {\em single-spin} bath, or just on the first two and the last two spins for a {\em two-spin} bath. For a single spin bath, 
we put $\hat{\cal L}_{\rm B}$ 
as part of $\hat{\cal L}_{\rm 1}$
as it commutes with all the other terms, so in the corresponding Suzuki-Trotter propagator
$\exp{(\hat{\cal L}_{\rm B} \tau)}$ factorizes out.
Our choice of $L_k$ will be such that $\hat{\cal L}_{\rm B}$ will have a single non-degenerate eigenvalue equal to $0$ with the corresponding eigenvector $\rho_B$ being a local equilibrium state, i.e. a direct product of thermal states of (each of) the edge spins, or the pairs of the edge spins. Furthermore, due to stability of completely positive dynamics, all other eigenvalues of
$\hat{\cal L}_{\rm B}$ should have negative real parts.\footnote{For an exact analysis of such open out-of-equilibrium quantum dynamics for Wigner-Jordan solvable models see Refs.~\cite{NJP:08,arxiv:08}.} 
  
 As a consequence, $\rho(t)$ will for sufficiently long time converge to NESS 
 $\rho_\infty = \lim_{t \to \infty}{\exp{(\hat{\cal L}t)}}\rho(0)$ of the entire spin chain, if starting from a generic initial state $\rho(0)$ having a non-zero overlap with $\rho_\infty$. Due to {\em ergodicity of the internal coherent dynamics} generated by the Hamiltonian $H$, we assume: (i) the resulting NESS $\rho_\infty$  will be unique, i.e. independent of the details of the initial condition
 $\rho(0)$, (ii) the resulting NESS $\rho_\infty$ will be 
(for {\em complex - say non-integrable and quantum chaotic - internal dynamics})  
 independent of the details of the bath operators $L_k$ - and will only depend on thermodynamic bath parameters, such as temperature, magnetization, chemical potential, etc. Both assumptions have been carefully checked and confirmed in the numerical simulations which are reported bellow.
 
Since non-unitary evolution does not preserve Schmidt decomposition structure of matrices $\mm{A}_i^{s_i}$ we apply local rotations every few steps in order to ``reorthogonalize'' matrices $\mm{A}_i^{s_i}$, recovering the Schmidt decomposition \cite{vidal}.
We note that Ref.\cite{zwolak} originally proposed the idea to use MPO for time-dependent solutions of Lindblad master equation, however it implemented it in a physically essentially different context of systems with bulk-dissipation where each spin has
been monitored by a Lindbladian bath. The fact that operator space entanglement is rather small in such a situation is not surprising, neither is the fact that it cannot describe coherent out-of-equilibrium many-body phenomena.

The goal of the present paper is to test the efficiency of MPO simulation of NESS as given by the asymptotic $t\to\infty$ solution of the
master equation driven only by the {\em boundary} Lindblad terms, for different (integrable and non-integrable) 1D spin chains,  checking along the way their transport behavior being for instance that of a normal (diffusive) conductor or displaying anomalous transport. 
One expects that quantum chaotic systems will display normal conduction \cite{Saito:03,Mejia:05}, that is, they will obey Fourier/Fick/Ohm's law in which the current $j$ is proportional to the gradient of a driving field $\varepsilon$, say local energy density/temperature (or spin density/magnetization, or
particle density/chemical potential, etc)
\begin{equation}
j=-\kappa \nabla \varepsilon.
\label{eq:fourier}
\end{equation}
For such normal conductors the transport properties should not depend on the details of the baths (assumption (ii) above). On the other hand, for the integrable models the choice of the baths could play a role. We are going to use two different models of the baths. In studies of spin conduction in the Heisenberg model we will use a single-spin bath while we are going to use a two-spin bath when studying energy transport in a quantum chaotic tilted Ising model. The reason for using a two-spin bath was in a better (faster) convergence to NESS with time $t$ and size $n$, which is probably due to the fact that the energy-density is a two-body operator, while the spin-density is a one-body operator.
 
\subsection{Single-spin bath}
When studying the spin (magnetic) transport we are going to couple the first and the last spin of the chain to a single-spin bath. 
For simplicity we shall bellow write only one part of $\hat{\cal L}_{\rm B}$ 
pertaining to a given edge spin, being either $l=1$ or $l=n$ and omitting the index $l$.
In all our numerical simulations with the single-spin bath we shall take the coupling strength $\gamma=1$.

There will be two Lindblad operators acting on the spin,
\begin{eqnarray}
L_1=\frac{1}{2}\sqrt{\Gamma_+}(\sigma^{\rm x}+\ii\, \sigma^{\rm y}),\qquad L_2=\frac{1}{2}\sqrt{\Gamma_-}(\sigma^{\rm x}-\ii\, \sigma^{\rm y}) 
\label{eq:1qubitL}\\
\Gamma_\pm=\sqrt{\frac{1\mp\tanh{\mu_{\rm L,R}}}{1\pm\tanh{\mu_{\rm L,R}}}}. \nonumber
\end{eqnarray}
Stationary state for $\hat{\cal L}_{\rm B}$ constructed from the above operators is $\rho_{\rm B}=\frac{1}{\Gamma_-+\Gamma_+} {\rm diag}(\Gamma_+,\Gamma_-)\propto \exp{(- \mu_{\rm L,R} \sigma^{\rm z})}$, 
with the average magnetization 
$\tr{(\rho_{\rm B}\sigma^{\rm z})}=-\tanh{\mu_{\rm L,R}}$. We stress that this is a stationary state of a single spin in the absence of Hamiltonian evolution. Parameters $\mu_{\rm L,R}$ of the left and right bath, respectively, play a role of an external thermodynamic potential enforcing a spin-density gradient, say a magnetization of a macroscopic magnet in contact with the edge spin.
Matrix representation of a super-propagator $\exp{(\hat{\cal L}_{\rm B}\tau)}$ in the Pauli basis $\sigma^\alpha$, $\alpha=0,1,2,3$, reads
\begin{equation}
\fl
\exp{(\hat{\cal L}_{\rm B}\tau)}=\pmatrix{
1 & 0 & 0 & 0  \cr
0 & {\rm e}^{-(\Gamma_++\Gamma_-)\tau} & 0 & 0  \cr
0 & 0 & {\rm e}^{-(\Gamma_++\Gamma_-)\tau} & 0  \cr
 \frac{\Gamma_+-\Gamma_-}{\Gamma_++\Gamma_-}\{ 1-{\rm e}^{-2(\Gamma_++\Gamma_-)\tau}\}  & 0 & 0 & {\rm e}^{-2(\Gamma_++\Gamma_-)\tau} \cr
}.
\label{eq:1L}
\end{equation}

\subsection{Two-spin bath}
Here we would like to construct a bath $\hat{\cal L}_{\rm B}$, 
i.e. determine the corresponding Lindblad operators $L_k$,  which produce a given unique stationary state $\rho_{\rm B}$ of a pair of spins,
e.g. such that $\rho_{\rm B}=\exp(-h/T)/\tr\exp(-h/T)$ is a local thermal (Gibbs) state with respect to some two-spin energy density operator $h=h(\sigma^{s_1}_1,\sigma^{s_2}_2)$.
In all our numerical simulations with the two-spin bath we shall take the coupling strength $\gamma=2$.

For the description of the methodology we shall assume that $\rho_{\rm B}$ is a known but completely general state given in terms of a $4\times 4$ matrix. We require that $\rho_{\rm B}$ is a unique eigenvector of $\hat{\cal L}_{\rm B}$ with the corresponding eigenvalue $0$, while all the other eigenvalues of $\hat{\cal L}_{\rm B}$ are negative. We are looking for a set of $L_k$ such that the resulting $\hat{\cal L}_{\rm B}$ will have the above properties. With the constraints so far the choice of $L_k$'s is not unique as we have fixed only one eigenvector and the corresponding eigenvalue of $\hat{\cal L}_{\rm B}$. So in addition we shall require that all other (negative) eigenvalues of $\hat{\cal L}_{\rm B}$ are equal to $-1$. Such $\hat{\cal L}_{\rm B}$ will produce the fastest possible convergence to $\rho_{\rm B}$ for a given fixed spectral norm of $\hat{\cal L}_{\rm B}$. 

Assuming first that $\rho_{\rm B}={\rm diag}(d_0,d_1,d_2,d_3)$ is diagonal
one can easily check that the following set of Lindblad operators $L_k$,
\begin{eqnarray}
L_{ij}= \sqrt{\frac{d_{m}}{32}}\, r^{i} \otimes r^{j},\qquad i,j=0,1,2,3, \label{eq:2qubitL}
\\
m=(i \hbox{ mod } 2)+2\,(j \hbox{ mod } 2), r^{i}=\{ \sigma^{\rm x}+\ii\, \sigma^{\rm y},\sigma^{\rm x}-\ii\, \sigma^{\rm y},\mathbbm{1} +\sigma^{\rm z},\mathbbm{1}-\sigma^{\rm z} \}, \nonumber
\end{eqnarray}
results in $\hat{\cal L}^{\rm diag}_{\rm B}$ satisfying the above conditions. Note that the above 16 Lindblad operators, labeled by a double index $k=(ij)$, can in fact be replaced by a set of 15 traceless operators leading to the same $\hat{\cal L}^{\rm diag}_{\rm B}$. In the Pauli basis the only nonzero matrix elements $({\cal L}^{\rm diag}_{\rm B})_{\alpha,\beta},\alpha,\beta\in\{0,\ldots,15\}$ are 
(assuming positive $d_i$ and $\tr{\rho_{\rm B}}=1$)
\begin{eqnarray}
(\hat{\cal L}^{\rm diag}_{\rm B})_{\alpha,\alpha}&=&-1,\qquad \alpha=1,\ldots,15 \cr
(\hat{\cal L}^{\rm diag}_{\rm B})_{15,0}&=&d_0-d_1-d_2+d_3, \cr
(\hat{\cal L}^{\rm diag}_{\rm B})_{12,0}&=&d_0+d_1-d_2-d_3, \cr
(\hat{\cal L}^{\rm diag}_{\rm B})_{3,0}&=&d_0-d_1+d_2-d_3. \label{eq:2qubitX}
\end{eqnarray}
Basis states are enumerated in such a way that the least significant bit is the first one, {\em i.e.} corresponding to the left factor in tensor products (\ref{eq:2qubitL}). 
Then, to obtain the bath data $L_k$ and $\hat{\cal L}_{\rm B}$ for a non-diagonal $\rho_{\rm B}$ we write the eigenvalue decomposition of $\rho_{\rm B}=V^\dagger d\, V$, where $d$ is diagonal and $V$ is unitary. In terms of the diagonal part $d$ we first obtain the Lindblad super-operator $\hat{\cal L}^{\rm diag}_{\rm B}$ as described above (\ref{eq:2qubitL},\ref{eq:2qubitX}) and then rotate it in the operator-space using the transformation $R$ induced by $V$. Writing the orthogonal matrix of $R$ in the Pauli basis $\sigma^{\vec{\alpha}}=\sigma^{\alpha_1}\otimes\sigma^{\alpha_2}$ we have
$R_{\vec{\alpha},\vec{\beta}}=\tr{(V^\dagger \sigma^{\vec{\alpha}}\, V \sigma^{\vec{\beta}})}/4$, giving the final Lindblad propagator
\begin{equation}
\exp{(\hat{\cal L}_{\rm B}\, \tau)}=R^T\exp{(\hat{\cal L}^{\rm diag}_{\rm B}\, \tau)} R.
\label{eq:2L}
\end{equation}
 
\section{Spin transport}

\label{sec:main1}

\begin{figure}[!h]
\centerline{\includegraphics{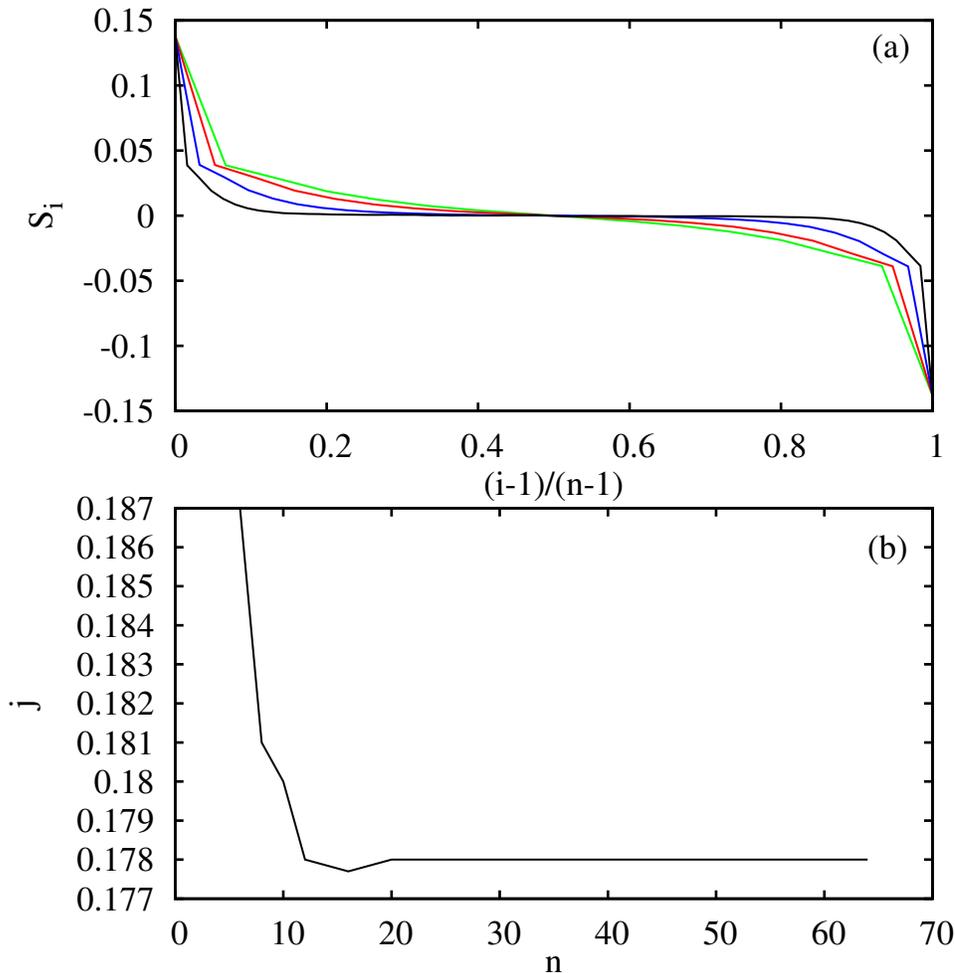}}
\caption{Spin profile (a) and spin current (b) for Heisenberg XXZ model at $\Delta=0.5$ and bath couplings (\ref{eq:1qubitL})
with $\mu_{\rm L,R}=\pm 0.22$. Data for different chain lengths $n=16,20,32,64$, from green (bright) to black curve, are shown. With increasing $n$ the spin profile gets flat -- an indication of an ideal spin conduction, visible also in the independence of the average spin current $j$ on the chain length $n$.}
\label{fig:h05profil}
\end{figure}

\begin{figure}[!h]
\centerline{\includegraphics{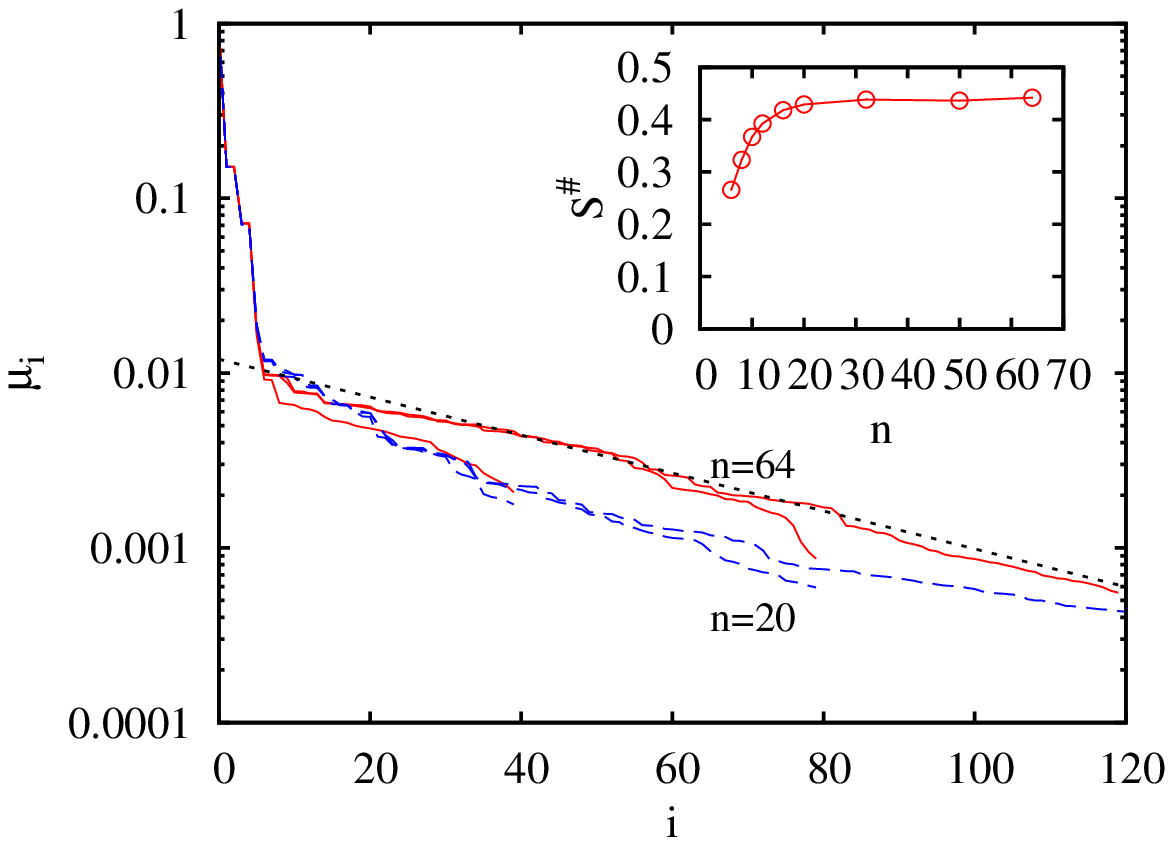}}
\caption{Spectrum of Schmidt coefficients $\mu_i$ for a symmetric bipartite splitting and different MPO dimensions $D=40,80,120$ of a NESS in the Heisenberg model (\ref{eq:heis}) with $\Delta=0.5$ coupled to two single-spin baths with $\mu_{\rm L,R}=\pm 0.22$. Full (red) curves are for the chain size $n=64$ while dashed (blue) curves are for $n=20$. Dotted line indicates the exponential decay $\exp(-i/40)$. Inset shows the dependence of OSEE $S^\sharp$ (\ref{eq:S}) on $n$. $S^\sharp$ does not grow appreciably with $n$, indicating the efficiency of MPO representation of NESS.}
\label{fig:h05S}
\end{figure}

Here we are going to study the spin transport in Heisenberg XXZ model with the Hamiltonian
\begin{equation}
H=\sum_{l=1}^{n-1} (\sigma_l^{\rm x} \sigma_{l+1}^{\rm x}+\sigma_l^{\rm y} \sigma_{l+1}^{\rm y}+\Delta \sigma_l^{\rm z} \sigma_{l+1}^{\rm z})+\sum_{l=1}^{n} h_l \sigma_{l}^{\rm z}.
\label{eq:heis}
\end{equation}
The first and the last spin will be coupled to a single-spin bath (\ref{eq:1L}). The initial state is chosen to be a product state 
$\rho(t=0)\propto \exp{(-\sum_l \mu_l \sigma_l^{\rm z})}$, where $\mu_l$ linearly interpolates between the left/right bath values $\mu_{\rm L,R}$.
We find that for times $t$ of the order of several times $n$ the state practically converges to NESS $\rho_\infty$, the properties of
which we are interested in. In particular, in order to asses the validity of the spin Fick's law we are going to calculate the {\em magnetization profile} 
(also referred to as the spin profile), $S_l=\tr{(\sigma_l^{\rm z} \rho_\infty)}$, and the local {\em spin current} defined as
\begin{equation}
j_l=\tr{\left[ (\sigma_l^{\rm x} \sigma_{l+1}^{\rm y}-\sigma_l^{\rm y} \sigma_{l+1}^{\rm x})\rho_\infty \right]}.
\label{eq:spinJ}
\end{equation}
We are going to study three different parameter regimes of the Heisenberg model corresponding to qualitatively different nature of many-body dynamics. The {\em integrable} XXZ Heisenberg model in the absence of magnetic field is known to display an ideal spin conduction for 
$\Delta < 1$ while it is probably a diffusive (normal) spin conductor for $\Delta > 1$~\cite{Narozhny:98,Zotos:99,Brenig:03,Long:03}.  
The last statement is quite controversial in the light of the algebraic integrability of the model \cite{Zotos:97} so it will be inspected more carefully in the present paper. As the third case we shall study spin transport in the XXZ model with staggered transverse magnetic field in the regime of quantum chaos, significantly improving numerical evidence for the conjecture (put forward for the case of heat transport in
 \cite{Saito:03,Mejia:05}) that quantum chaos corresponds to normal diffusive transport (Fourier's, Ohm's or Fick's law).

\subsection{Ideally conducting regime, $\Delta=0.5$}
We will first take $\Delta=0.5$ and no magnetic field, $h_l=0$, in order to test the method in a regime where Heisenberg model is an ideal spin conductor (having a non-vanishing Drude weight at any, say infinite temperature \cite{Zotos:99}). The left and the right bath parameters are set to $\mu_{\rm L}=0.22$ and $\mu_{\rm R}=-0.22$. The results of numerical simulation are shown in figure~\ref{fig:h05profil}. One can clearly see that the system is an ideal conductor, that is, it does not obey spin Fick's law because the spin current is found proportional to the magnetization difference and {\em not} its gradient. This is a typical property of completely integrable systems.

Considering numerical efficiency of MPO simulation of NESS, we show in figure~\ref{fig:h05S} the Schmidt spectrum for a symmetric bipartite cut of NESS and different MPO dimensions $D$. Numerical data suggest that the tails of the spectra of Schmidt coefficients decay {\em exponentially}. OSEE $S^\sharp$ (\ref{eq:S}) does not seem to grow with $n$ (although based on numerics we can not exclude a very slow growth) suggesting the method is efficient, i.e. computation time is asymptotically polynomial in $n$ (linear in $n$ if OSEE converges). 

\subsection{Normal conductor, $\Delta=1.5$}
\begin{figure}[!h]
\centerline{\includegraphics{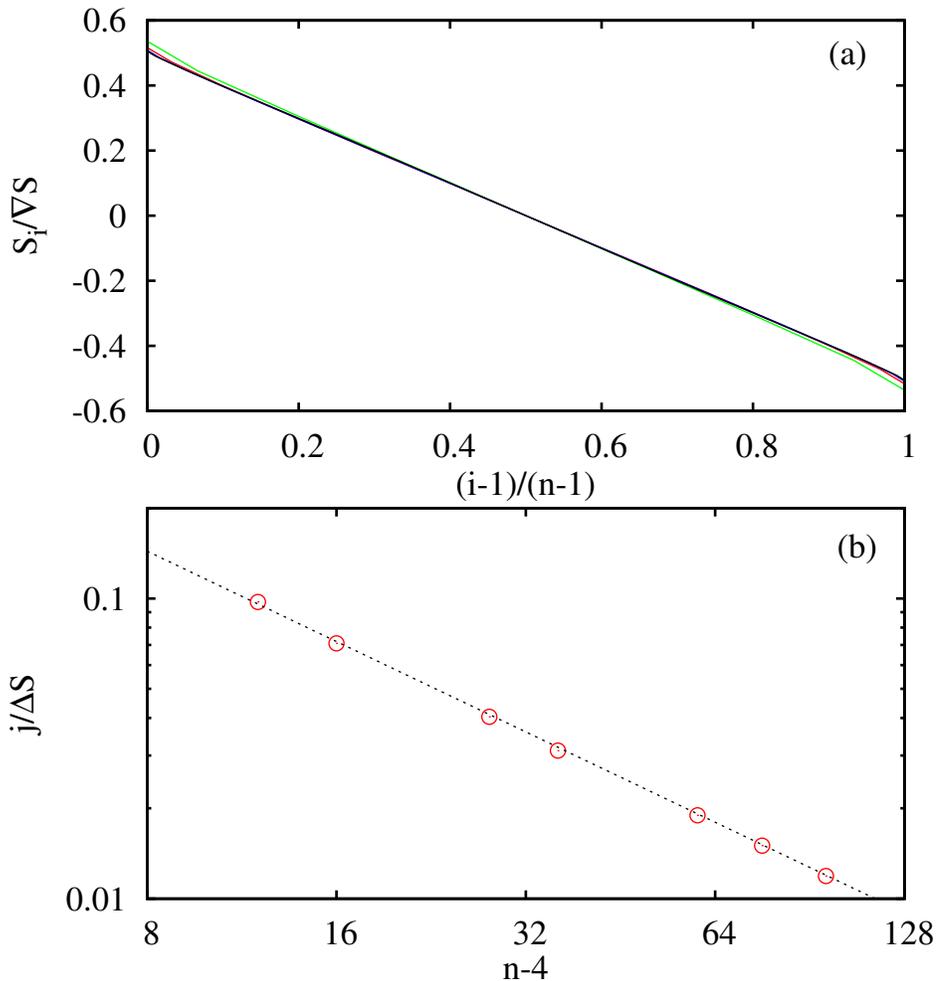}}
\caption{Spin profiles and current versus system size in NESS for the Heisenberg XXZ model at $\Delta=1.5$ and the bath coupling parameters $\mu_{\rm L,R} = \pm 0.02$. Data for different chain lengths $n=16,32,64,100$, from green (bright) to black curve, 
almost overlap (frame (a)). Apart from the boundary effects the spin profile is linear, suggesting normal conduction and diffusive transport. This is confirmed in frame (b), where we show the dependence of the scaled spin current $j/\Delta S$ on the chain length $n$. Dotted line is $j/\Delta S \sim 1.15/(n-4)$, indicating spin ``Fick's law" with the spin conductivity $\kappa=1.15$.}
\label{fig:h15profil}
\end{figure}

Increasing $\Delta$ above $1$ the Heisenberg model becomes a normal (diffusive) spin conductor (but {\em not} a diffusive heat conductor!) despite its integrability. We take $\Delta=1.5$ in the absence of the magnetic field, $h_l=0$, and a single-spin bath with a weak driving 
$\mu_{\rm L,R}=\pm 0.02$ (\ref{eq:1L}) in order to make sure that non-linear transport features do not obscure the effect. Figure~\ref{fig:h15profil} clearly demonstrates that the system is indeed normal spin conductor. Spin profiles in NESS are, apart from boundary effects, perfectly linear. To reduce the boundary effects we drop the leftmost and the rightmost two spins when calculating the drop of magnetization $\Delta S = S_{n-2} - S_{3}$, and its gradient 
$\nabla S = \Delta S/(n-4)$.
Spin current is clearly proportional to the gradient of magnetization $\nabla S$, or at fixed bath data, to $\sim 1/n$, see figure~\ref{fig:h15profil}b. Note that the largest chain size $n=100$ is much larger than what has been numerically achievable with other methods, like various Monte Carlo wave-function methods \cite{Mejia:05,Mejia:07,Wichterich:07}. 

\begin{figure}[!h]
\centerline{\includegraphics{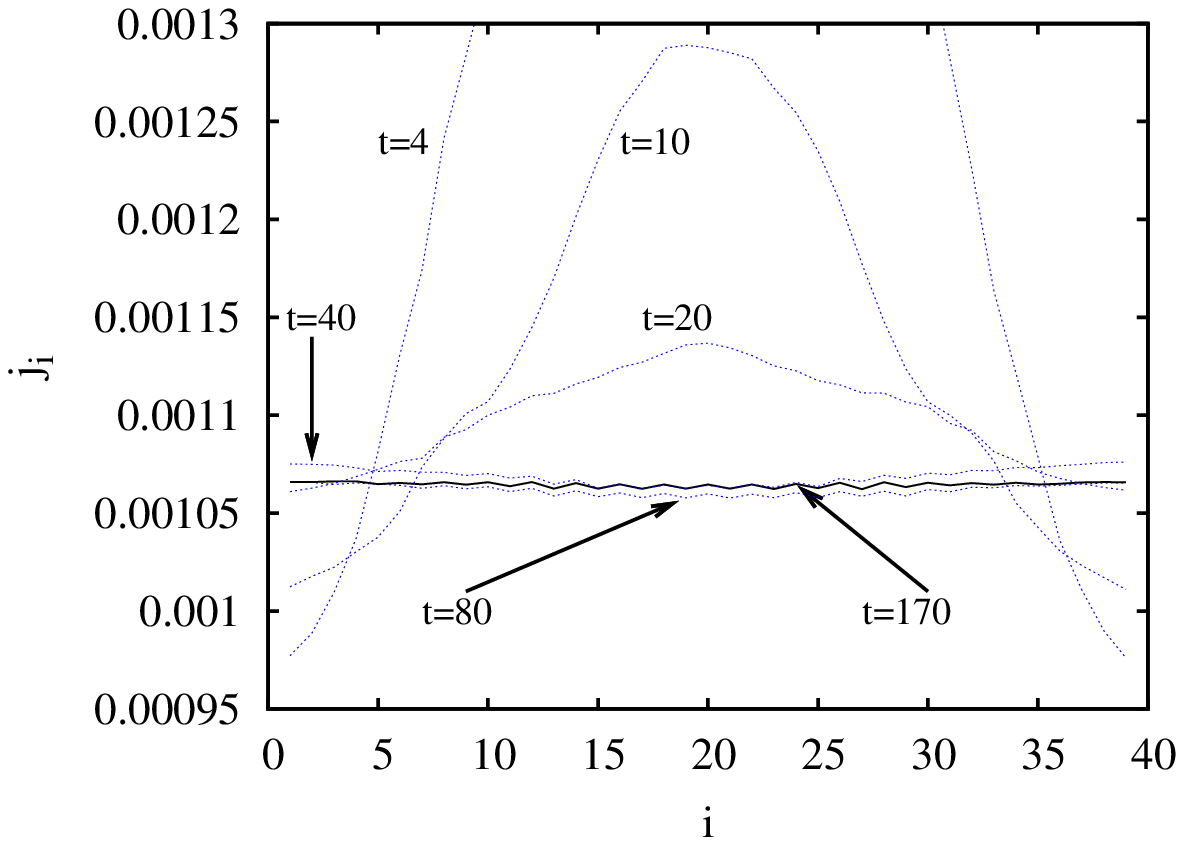}}
\caption{Convergence of the spin current profiles (dependence of the local current $j_i$ on the lattice site $i$) with time
for the Heisenberg XXZ model at $\Delta=1.5$ (the case of Fig.~\ref{fig:h15profil}) and for size $n=40$. 
Snapshots at times $t=4,10,20,40,80,170$ are shown, for, respectively, automatically adapted
MPO dimensions $D=40,60$ and later $D=80$.}
\label{fig:h15j}
\end{figure}

To show the convergence of $\rho(t)$ to the asymptotic NESS $\rho_\infty$ we show in figure~\ref{fig:h15j} the snapshots at different instants of time $t$ of local spin current profiles $j_i$ (\ref{eq:spinJ}). Clearly, a tendency towards a uniform current profile at NESS required by continuity equation is observed. 

The tail of the Schmidt spectrum decays here qualitatively slower than for an ideally conducting case (e.g. $\Delta=0.5$ of fig.~\ref{fig:h05S}b), exhibiting perhaps an asymptotic power law decay $\mu_i \sim 1/i^{1.25}$. 
Note that if Schmidt coefficients decay algebraically as $\mu_i \sim 1/i^p$ with some power $p$, then the OSEE $S^\sharp$ converges to its exact value for $D \to \infty$ as $S^\sharp_{D=\infty}-S^\sharp_D \sim \log_2{(D)}/D^{2p-1}$. For our $p \approx 1.25$ this would mean a very slow $\sim \log_2{(D)}/D^{1.5}$ convergence of $S^\sharp$. Since we are limited to relatively small dimensions, of order $D \sim 100$, we are here not able to asses whether $S^\sharp$ saturates with $n$ or not.

\begin{figure}[!h]
\centerline{\includegraphics{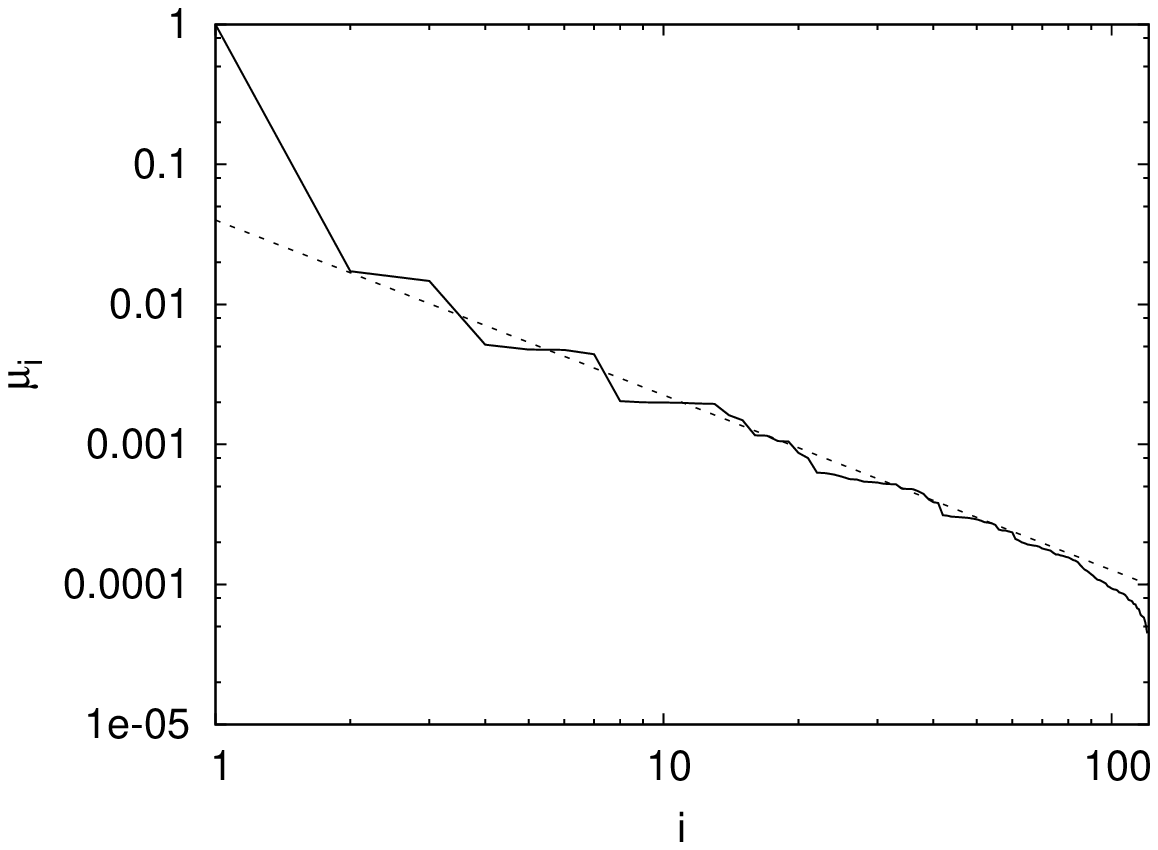}}
\caption{Spectrum of Schmidt coefficients $\mu_i$ of NESS (at convergence time $t=250$) for a symmetric bipartite cut of the Heisenberg XXZ model at $\Delta=1.5$. MPO Dimension is $D=120$ and the spin chain length $n=64$. Dotted line indicates a power law decay
$i^{-1.25}$, with finite size effects setting in when $i\approx D$. }
\label{fig:h15l}
\end{figure}

\subsection{Normal conductor, $\Delta=0.5$ and staggered transverse field}

\begin{figure}[!h]
\centerline{\includegraphics{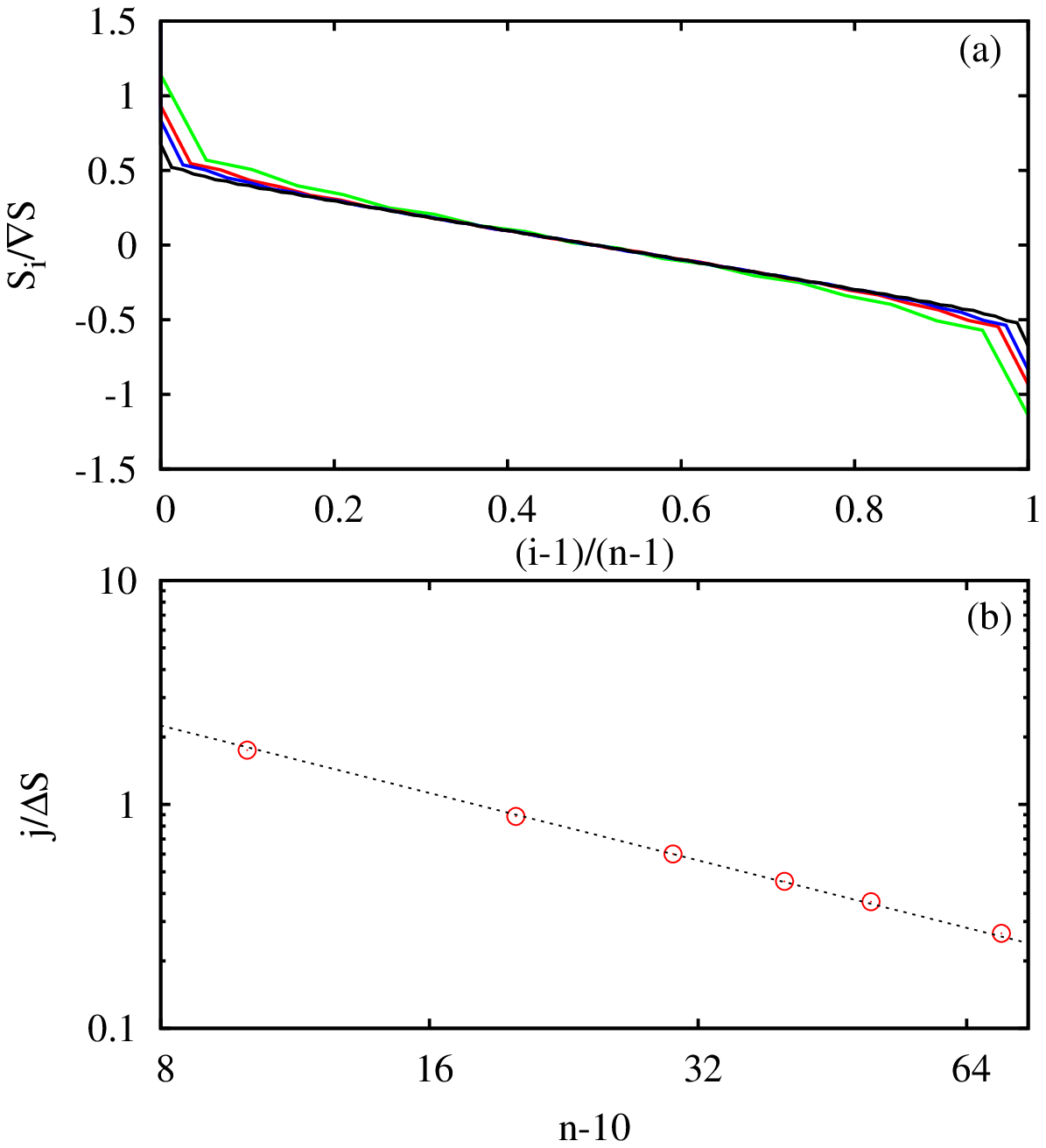}}
\caption{Spin profile (frame (a)) for quantum chaotic Heisenberg XXZ model  at $\Delta=0.5$ in a staggered transverse field 
$h_l = (0,-0.5,0,-0.5,\ldots)$. Data for chain lengths $n=20,30,40,80$,  from green (bright) to black curve, are shown. Apart from the boundary effects the spin profile is linear, suggesting normal diffusive spin conduction. This is confirmed in frame (b), where we show the dependence of the scaled average spin current $j/\Delta S$ on $n$. Dotted line is $j/\Delta S \sim 18.0/(n-10)$, indicating ``Fick's law" with the spin conductivity $\kappa=18.0$.}
\label{fig:h05cprofil}
\end{figure}

Here we are going to take the Heisenberg model (\ref{eq:heis}) with $\Delta=0.5$ and staggered transverse magnetic field,
$h_{2l+1}=0$ and $h_{2l}=-1/2$. We use the bath parameters $\mu_{\rm L,R}=\pm 0.1$. 
We have checked that for these parameter values the spectrum of the Hamiltonian (\ref{eq:heis}) exhibits the characteristics of {\em quantum chaos}, i.e. the energy level spacing distribution agrees with universal predictions of random matrix theory with no free parameters,
so the spin conduction is expected to be normally diffusive and to obey the Fick's law.
The results shown in Fig.\ref{fig:h05cprofil}  clearly confirm this expectation, namely the spin density profiles are linear, and the spin current decays as $\propto 1/n$.
Note that this figure is very similar to Fig.\ref{fig:h15profil} for the diffusive-integrable case with $\Delta=1.5$,
a subtle difference may be that now the boundary effects seem to be stronger.
We have thus dropped border 5 spins at each end of a chain when calculating the spin drop $\Delta S = S_{n-5} - S_{6}$ and the gradient $\nabla S = \Delta S/(n-10)$. 

\begin{figure}[!h]
\centerline{\includegraphics{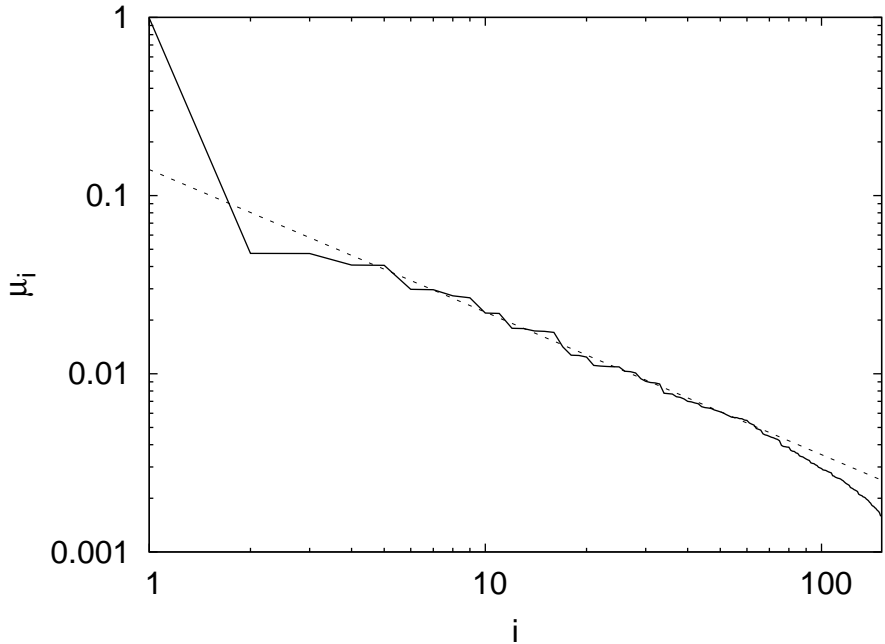}}
\caption{Spectrum of Schmidt coefficients $\mu_i$ of NESS (at convergence time $t=200$) for a symmetric bipartite cut of 
quantum chaotic Heisenberg model in a staggered field.  MPO Dimension is $D=150$ and size $n=50$. Dotted line indicates a power law decay
$i^{-0.80}$.}
\label{fig:h05sch}
\end{figure}

The tail of the Schmidt spectrum shown in figure~\ref{fig:h05sch} seems to decay even slower now, 
$\mu_i \sim 1/i^{0.80}$, making the simulation harder than in both integrable cases without the magnetic field.
This is consistent with the previous observations in MPO simulations of long-time evolution in the Heisenberg picture \cite{PRE:07}.

\section{Energy transport}

\label{sec:main2}

As the last test we are going to study the heat conductivity in Ising model placed in a tilted magnetic field, {\em tilted Ising model} for short, 
with the Hamiltonian
\begin{equation}
\fl
H=\sum_{l=1}^{n-1} h_{l,l+1},\qquad h_{l,l+1}=-2\sigma_l^{\rm z} \sigma_{l+1}^{\rm z}+\frac{1}{2}(h_{\rm x}\sigma_l^{\rm x}+h_{\rm z}\sigma_l^{\rm z})+\frac{1}{2}(h_{\rm x}\sigma_{l+1}^{\rm x}+h_{\rm z}\sigma_{l+1}^{\rm z}),
\label{eq:TI}
\end{equation}
with $h_{\rm x}=3.375$ and $h_{\rm z}=2$. For these parameters the system is quantum chaotic, see {\em e.g.}~\cite{Mejia:05}, and displays normal heat conduction~\cite{Mejia:05,Mejia:07}, however, previous simulations were limited to system sizes $n\le 20$. 
For studies of heat transport in other quantum chaotic systems see also~\cite{Saito:03,Steinigeweg:06}. Local energy current is in this case
\begin{equation}
j_l=2 h_{\rm x} \tr{\left[ (\sigma^{\rm z}_{l-1} \sigma^{\rm y}_l-\sigma_l^{\rm y} \sigma_{l+1}^{\rm z}) \rho_\infty \right]}.
\label{eq:je}
\end{equation}
When we used the single-spin bath (\ref{eq:1L}) the convergence of local energy current to a homogeneous site-independent value expected for NESS was rather slow, meaning that the spectral gap of the quantum Liouville operator $\hat{\cal L}$ is inconveniently small
for such a bath model. Therefore, we rather used a two-spin bath (\ref{eq:2L}) for which these effects are smaller. 
We set the temperature of the left bath to $T_{\rm L}=20$ and of the right bath to $T_{\rm R}=30$. 

Since in an out-of-equilibrium system the definition of local temperature may not be completely unambiguous we have instead of looking at the temperature profile, computed the energy density profile $\varepsilon_l = \tr \rho_{\infty} h_{l,l+1}$. As discussed in \cite{Mejia:05}, the energy density uniquely determines the temperature in the equilibrium (thermal) state and it can also be used as a good measure of local temperature out of (but not too far from) equilibrium. 

\begin{figure}[!h]
\centerline{\includegraphics{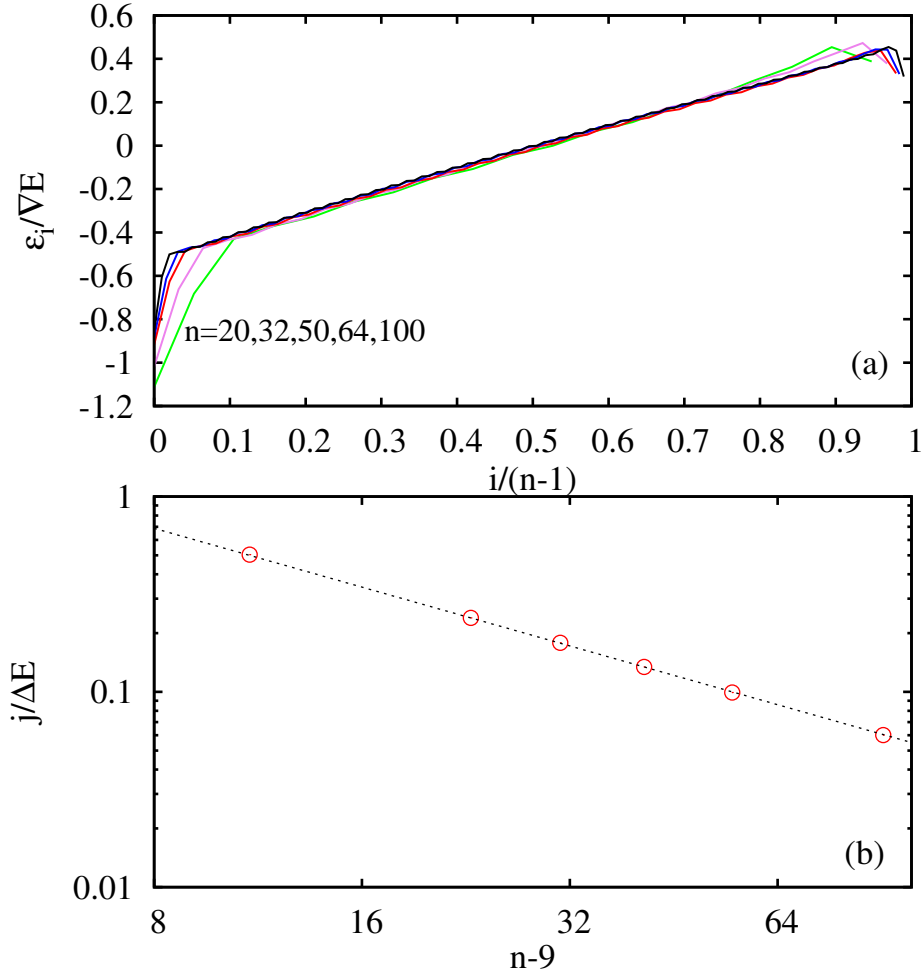}}
\caption{Local energy density profiles in NESS for quantum chaotic tilted Ising model (\ref{eq:TI}). Data for chain lengths $n=20,32,50,64,100$,  from green (bright) to black curve, almost overlap (frame (a)). Apart from the boundary effects the energy density profile is linear, suggesting normal heat conduction. This is confirmed inframe (b), where we show the dependence of the scaled energy current $j/\Delta E$ on $n$. Dotted line is $j/\Delta E \sim 5.5/(n-9)$, indicating the Fourier law behavior with the conductivity $\kappa=5.5$.}
\label{fig:tiprofil}
\end{figure}

The boundary effects seen in figure~\ref{fig:tiprofil} are again relatively strong. Because of these, and because we kept $T_{\rm L,R}$ - 
which is a ``non-interacting temperature'' of a 2 spin system - fixed for all $n$, the actual local temperature varies slightly with $n$. In figure~\ref{fig:tiprofil}a we therefore appropriately shifted individual energy profiles in vertical direction in order to get scaled overlapping curves. 
To decrease finite-size effects at boundaries we dropped border 4 spins when calculating the energy difference $\Delta E$ or the energy density gradient $\nabla E = \Delta E/(n-9)$ 
(fig.~\ref{fig:tiprofil}).

Taking all these into account, we have confirmed the Fourier's law of heat conduction $j = \kappa \Delta E/(n-9)$, where $\Delta E$ was the energy density difference between spins $(5,6)$ and $(n-5,n-4)$, with an excellent numerical accuracy for $10\le n \le 100$.

We have also looked at the tail of the spectrum of the Schmidt coefficients which again exhibits power law decay, best fitted with
$\mu_i \propto i^{-0.72}$.

\begin{figure}[!h]
\centerline{\includegraphics{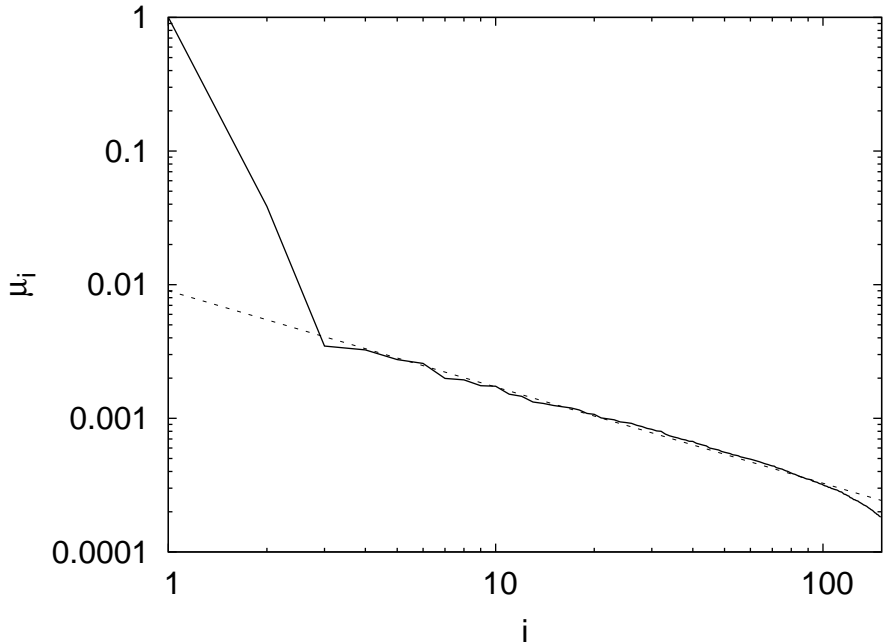}}
\caption{Spectrum of Schmidt coefficients $\mu_i$ of NESS (at convergence time $t=500$) for a symmetric bipartite cut of 
quantum chaotic titled Ising model (for the case of fig.~\ref{fig:tiprofil}). MPO dimension is $D=150$ and size $n=50$.
Dotted line indicates a power law decay
$i^{-0.72}$.
}
\label{fig:til}
\end{figure}

\section{Summary and discussion}

\label{sec:disc}

In the present paper we have demonstrated numerical simulation of NESS for strongly but locally interacting open quantum systems in 1D
in terms of the MPO ansatz. We have considered the most difficult case where dynamics in the bulk is fully coherent (Hamiltonian), 
and dissipation (coupling to the baths) only takes place at the boundary (ends of the chain). In this setting we have been able to confirm
the laws of diffusive transport, such as Fourier's law of heat conduction and (spin) Fick's law for spin conduction, in the cases where the 
underlying model is strongly non-integrable and displays the features of quantum chaos, or even when it is integrable but all the conserved quantities are irrelevant for the transporting current, like in the case of spin-conduction in Ising-like Heisenberg XXZ chain. 
The main purpose of our paper was to demonstrate that such simulations were now possible for system sizes of order of $n=100$ spins 1/2, which is considerably larger than with the competing methods (usually based on Monte Carlo wave-function techniques, where presently only $n\approx 20$ is reachable for fully out-of-equilibrium simulations).

As for the quantitative analysis of the efficiency of MPO ansatz for NESS we have analyzed the spectrum of the Schmidt decomposition for the worst-case
(half-half) bipartition of the chain, when treating the density matrix of NESS as an element of the Hilbert space of operators.
Summarizing these results, we have found that {\em completely integrable} and {\em ideally conducting cases}, in our example $XY$-like Heisenberg XXZ chain ($|\Delta| < 1 $), are easiest to simulate since there the tail of the  spectrum of Schmidt coefficients decays exponentially so MPO of a rather modest matrix dimension gives a good description of NESS.

For still integrable systems, but for which all the conservation laws are irrelevant to the transporting current in NESS, such as the Ising-like Heisenberg XXZ chain ($|\Delta| > 1$), we have found qualitatively worse efficiency, namely there the tails of the Schmidt spectrum appear to exhibit power law tails $\mu_i \propto i^{-1.25}$.

At last, for non-integrable systems in the regime of quantum-chaos, the efficiency of simulation appears again worse than in the integrable cases above (in our examples we have looked into the Heisenberg XXZ chain in a staggered transverse field, and Ising spin chain in a tilted magnetic field). Namely, there we have found algebraically decaying tails of the Schmidt spectrum $\mu_i \propto i^{-p}$ with the power $p \in [0.7,0.8]$.

Based on our results we also conclude that zero-temperature (ground state) properties of the 
system, whether being critical or gaped, do not influence far-from-equlibrium properties of NESS.

The work has been financially supported by Program P1-0044 and Grant J1-7347 of Slovenian Research Agency (ARRS).

\end{document}